\documentclass[useAMS,usenatbib]{mn2e}
\usepackage{myaasmacros}
\usepackage{graphicx}
\usepackage{ulem}

\def\ltsima{$\; \buildrel < \over \sim \;$}
\def\simlt{\lower.5ex\hbox{\ltsima}}   
\def\gtsima{$\; \buildrel > \over \sim \;$}
\def\simgt{\lower.5ex\hbox{\gtsima}}

\bibliography{../../BibTeX/refs}

\title[MAGIC II]{THE MAGELLANIC INTER-CLOUD PROJECT (MAGIC) II: SLICING UP THE BRIDGE}
   
\author[No\"el et al.]{N. E. D. No\"{e}l$^{1}$\thanks{E-mail: n.noel@surrey.ac.uk}, B. C. Conn$^{2}$, J. I. Read$^{1}$,  R. Carrera$^{3,}$$^{4}$, A. Dolphin$^{5}$, and H.-W. Rix$^{6}$
\\
$^{1}$Department of Physics, University of Surrey, Guildford, GU2 7XH, UK\\
$^{2}$Gemini Observatory, Recinto AURA, Colina El Pino s/n, La Serena, Chile\\
$^{3}$Instituto de Astrof{\'i}sica de Canarias, C/ V{\'i}a L\'actea s/n, 38200, La Laguna, Tenerife, Spain\\
$^{4}$Departamento de Astrof{\'i}sica, Universidad de La Laguna, Tenerife, Spain\\
$^{5}$Raytheon Company, PO Box 11337, Tucson, AZ, 85734-1337, United States of America\\
$^{6}$Max Planck Institut f\"{u}r Astronomie, K\"{o}nigstuhl 17, 69117, Heidelberg, Germany}

\date{Released 2002 Xxxxx XX}

\pagerange{\pageref{firstpage}--\pageref{lastpage}} \pubyear{2002}

\def\LaTeX{L\kern-.36em\raise.3ex\hbox{a}\kern-.15em
    T\kern-.1667em\lower.7ex\hbox{E}\kern-.125emX}

\begin{document}

\label{firstpage}

\maketitle

\begin{abstract}

The origin of the gas in between the Magellanic Clouds (MCs), known as the Magellanic Bridge (MB), has always been the subject of controversy. To shed light into this, we present the results from the MAGIC II project aimed at probing the stellar populations in ten large fields located perpendicular to the main ridge-line of HI in the Inter-Cloud region. We secured these observations of the stellar populations in between the MCs using the WFI camera on the 2.2 m telescope in La Silla.
Using colour-magnitude diagrams (CMDs), we trace stellar populations across the Inter-Cloud region. In good agreement with MAGIC I, we find significant intermediate-age stars in the Inter-Cloud region as well as young stars of a similar age to the last pericentre passage in between the MCs ($\sim$200 Myr ago). We show here that the young, intermediate-age and old stars have distinct spatial distributions. The young stars correlate well with the HI gas suggesting that they were either recently stripped from the SMC or formed in-situ. The bulk of intermediate-age stars are located mainly in the bridge region where the HI column density is higher, but they are more spread out than the young stars. They have very similar properties to stars located $\sim$ 2 Kpc from the SMC centre, suggesting that they were tidally stripped from this region. Finally, the old stars extend to some 8 Kpc from the SMC supporting the idea that all galaxies have a large extended metal poor stellar halo.

\end{abstract}


\section{Introduction}

A key challenge in modern astrophysics is to understand how galaxies form and evolve. In particular, interactions between galaxies are an important driver of galaxy formation and evolution (e.g. \citealt{1978ApJ...219...46L}, \citealt{1992Natur.360..715B}). These encounters reshape galaxies by transferring mass, energy, gas and metals between them, creating new sites of star formation both within the galaxies and in stripped gas amidst them.
 Important clues about the galaxy formation process lie in the faint edges of galaxies.
 Galaxy formation simulations suggest that almost all galaxies, even the smallest, should contain an extended old, metal poor, stellar halo that is primarily composed of the detritus from galactic mergers 
 (e.g. \citealt{1978MNRAS.183..341W}; \citealt{1978ApJ...225..357S}; \citealt{2005ApJ...635..931B}; \citealt{2006MNRAS.366..429R}; \citealt{2010MNRAS.406..744C}). 
 This is supported by a host of observations from large spirals (e.g. \citealt{1954AJ.....59..307R}; \citealt{1962ApJ...136..748E}; \citealt{2006ApJ...648..389K}; \citealt{2006ApJ...647L..25M}; 
 \citealt{2006ApJ...653..255C}; \citealt{2012ApJ...760...76G}; \citealt{2013MNRAS.431.1121T}; \citealt{2013ApJ...766..106M}; \citealt{2014ApJ...787...30D}; \citealt{2014A&A...562A..73G}) to dwarf galaxies (e.g. \citealt{2000AJ....119..177A}; \citealt{2003Sci...301.1508M}; \citealt{2007ApJ...665L..23N}; \citealt{2009ApJ...704.1730P}; \citealt{2012AJ....143...52S}). 
 Such halos retain a fossil record of past mergers and interactions in the form of extended tidal debris (e.g. \citealt{1978ApJ...225..357S};  \citealt{2001ApJ...549L..63M}; \citealt{2006ApJ...649..201M}; \citealt{2007ApJ...663..960S}; \citealt{2014Natur.507..335A}). 
 
 The distribution and extent of such debris, or a characteristic underlying old stellar halo are both sensitive to the total mass and size of a galaxy
 (e.g. \citealt{2007ApJ...665L..23N}; \citealt{2013ApJ...768..109N}; \citealt{2014ApJ...787...30D}). Thus, probing the faint edges of
galaxies can also give us constraints on their dark matter content. Observing these remote edges of galaxies is a complicated task due to the faintness of these outlying regions, 
typically $\mu$$\ge$28 mag/arcsec$^{2}$ (e.g. \citealt{2007ApJ...665L..23N}). Reliable surface brightness measurements at these faint levels require accurate flat field division, sky determination, bright star masking, among others (e.g. \citealt{2009AJ....138.1469B}). 

 Our nearest irregular neighbours, the Magellanic Clouds (MCs) constitute a unique opportunity to study a gas rich interacting system close-up and to investigate faint outskirts of galaxies. First, their close proximity ($\sim $50-70\,Kpc away) enables us to resolve their individual stars, which allows to directly probe fainter surface brightness levels placing constraints on ages and metallicities without the need to use stellar population synthesis (e.g. \citealt{2013ApJ...772...58N}). Second, due to their mutual association over the past few Gyrs (as determined from their proper motions: \citealt{2006ApJ...652.1213K}, \citealt{2007ApJ...668..949B}, \citealt{2008AJ....135.1024P}, \citealt{2011AJ....141..136C}, \citealt{2012MNRAS.421.2109B}, \citealt{2013ApJ...764..161K}, \citealt{2014ApJ...781..121V}, \citealt{2014A&A...562A..32C}; and numerical simulations: \citealt{2011MNRAS.416.2359B}; \citealt{2012ApJ...750...36D}; \citealt{2013MNRAS.428.2342B}). Third, because of their significant gas reservoirs (e.g. \citealt{2000MNRAS.315..791S}, \citealt{2010PASP..122..683K}, \citealt{2011ApJ...741...12B}, \citealt{2013MNRAS.429.2527M}).
Indeed various gaseous features were created by the interactions of the MCs with each other and with the Milky Way (MW) such as the Magellanic Bridge (MB; \citealt{1963AuJPh..16..570H}), 
 a common envelope of HI spanning $\sim$13 Kpc in which both the Large Magellanic Cloud (LMC) and the Small Magellanic Cloud (SMC) are embedded; the Magellanic Stream, a pure gaseous stream trailing the MCs as they orbit the MW (\citealt{1974ApJ...190..291M}, \citealt{2010ApJ...723.1618N}) connecting both of the MCs with the MW, and the Leading Arm, a mass of high-velocity clouds leading the MCs (\citealt{2000PASA...17..1P}). 
 
Since its discovery, the MB was believed to have a tidal origin, forming during an encounter between the MCs \citep{1985PASAu...6..104M}. 
The SMC Wing --the Eastern protuberance of the SMC main body-- extends towards the LMC into the Bridge \citep{1990AJ.....99..191I} indicating that 
this tidal interaction could have formed a disk around the SMC that was later on stripped to form the MB \citep{1992AJ....103.1234G}.
Part of the strong observational evidence of the tidal stripping scenario includes: 
\begin{itemize}

\item Carbon stars also in the MB area (e.g. \citealt{2000AJ....119.2789K}).

\item Intermediate-age stellar populations in the Southern  SMC \citep{2007ApJ...665L..23N}.

\item Kinematically peculiar stars in the LMC that might be evidence of being captured from the SMC  \citep{2011ApJ...737...29O}.
 
 \item Possible evidence of extra-tidal stars at $\sim$10.6$^{\circ}$ from the SMC centre \citep{2011ApJ...733L..10N}.
 
 \item Old stars found in the MB  \citep{2013A&A...551A..78B}.
 
 \item  The red clump stars found in the OGLE IV fields\footnote{http://ogle.astrouw.edu.pl/} \citep{2014ApJ...795..108S}.
 
 \item The stellar structure in the eastern SMC identified by \cite{2013ApJ...779..145N}.  These authors suggest that this is a stellar counterpart of the HI MB 
that was tidally stripped from the SMC around 200 Myr ago (during a close encounter with the LMC). This implies an increment in the self-lensing of the SMC stars, in line with the 
 tidal origin for the microlensing events reported towards the LMC by \cite{2013MNRAS.428.2342B}.  
 
 \item The intermediate-age stars in the MAGIC I \citep{2013ApJ...768..109N} in a field closer to the SMC make up 28\% of all stars in that region. These
  are not present in fields in a direction pointing away from the LMC. This provides potential evidence that these intermediate-age stars could have been tidally stripped from the SMC.
  
 \end{itemize}

 With the advent of the new STEP\footnote{STEP: The SMC in Time: Evolution of a Prototype interacting late-type dwarf galaxy'} survey \citep{2014arXiv1405.1028R} that will cover the main body of the SMC (32 deg$^2$), the MB (30 deg$^2$) and a small part of the Magellanic Stream (2 deg$^2$), 
 evidence will likely continue to mount.

In spite of the above striking evidence, there still remain some results at odds with the tidal scenario. \cite{1998AJ....115..154D}  carried out the first extensive study of the Inter-Cloud stellar population finding only young stars ($\sim$10-25 Myr old). A similar result was also found by \cite{2007ApJ...658..345H} who studied twelve fields in the MB  finding only old stars east of $\alpha$$\sim$03h 18' and $\delta$$\sim$-74$^{\circ}$ and only young stars in their Inter-Cloud fields along the MB's main ridge-line, with no evidence of an intermediate-age component. 
As they did not find evidence of tidally stripped stars, \cite{2007ApJ...658..345H} concluded that the material stripped from the MCs into the MB must be almost purely gaseous, while any young stars must have formed in-situ in the MB.

With the purpose of shedding light on the stellar content of the MCs' Inter-Cloud area, we have initiated the MAGIC (MAGellanic Inter-Cloud) project, aimed at disentangling the population age and distribution in the Inter-Cloud region as well as probing the faint outskirts of the MCs in the direction of the Bridge. 
As stated above, in MAGIC I \citep{2013ApJ...768..109N} we performed a quantitative Color-Magnitude Diagram (CMD) fitting analysis of two large Inter-Cloud fields, finding that some $\sim$28\% of the stars in one of the fields have ages between $\sim$1 and 10 Gyr old (intermediate-age population).  \cite{2007ApJ...658..345H}
 likely missed this intermediate-age population because their CMDs were substantially shallower than the MAGIC CMDs that reach the oldest MS turnoffs, of key importance for spotting intermediate-age stars (see e.g. \citealt{2008PASP..120.1355N}).
The results from MAGIC I provide more evidence that these intermediate-age stars could have been stripped from the SMC, supporting the tidal scenario.

In the present paper we focus on the stellar content of ten fields strategically located perpendicular to the MAGIC I field where intermediate-age population was found represented by a purple triangle in Figure \ref{fig_bridge}
(field B2, located at $\sim$8 Kpc from the SMC centre in the Inter-Cloud region). Five of the fields are radially away from the MB towards the North and five are radially away from the MB towards the South of field B2 in MAGIC I.  
This arrangement allows us to differentially compare stellar populations inside the bridge region with directly neighbouring fields, allowing us to separate distant SMC stars from those possibly stripped from the SMC's interior. Our goal here is to probe the outskirts of the SMC in the MB direction.

This paper is organized as follows. In section \ref{data}, we describe the observations, data reduction and photometry. The analysis and results are discussed in section \ref{results}. Finally, in section \ref{discussion} we discuss our results and present our conclusions.

\section{Observations, Data Reduction and Photometry}\label{data}

 Following our experience with previous observations for MAGIC I \citep{2013ApJ...768..109N},  we obtained B and R band images of ten fields between the nights of 18$^{\rm th}$ to 26$^{\rm th}$ December 2011 using the Wide Field Imager (WFI) on the 2.2m telescope in La Silla.  The WFI consists of 8 chips covering a field of view of approximately 30$\arcmin \times$30$\arcmin$ and has a pixel scale of 0.24$\arcsec$/pixel. Each B-band exposure was 1200 seconds with 900 seconds for the R-band. An overview of the field positions can be seen in Figure~\ref{fig_bridge} with the MAGIC II fields depicted as squares showing the Northern and Southern extensions from our base position. The Northern arm extends radially away from the SMC, while the Southern arm extends perpendicular from the Magellanic Bridge (we will return to this Figure later). In Table~\ref{ObsTable}, we present the field name, RA/DEC centres of the observed fields, observation date, seeing estimate, filter, airmass, and the 50\% completeness level of the data. N represent the fields located North of B2 and S those located South of B2, the increasing numbers in S (S1-S5) represent fields of increasing Declination from the bridge ridge-line while the increasing numbers in N (N1-N5) represent fields of decreasing Declination. The overlaps between the MAGIC I field (B2) and the first Northern field is 16.8 arcminutes and 14.3 arcminutes for the Southern field. Each subsequent field has an overlap of approximately 2-3 arcminutes.
 For the overlap with MAGIC I this corresponds to at 2300 stars and for the overlaps between fields we obtain an average of 780 stars to determine the photometric offsets.

The data reduction steps performed here followed the same procedure as in MAGIC I. The basic reduction was performed by the Cambridge Astronomical Survey Unit pipeline (CASU; \citealt{2001NewAR..45..105I}) with nightly master bias and master flat frames from adjacent nights were combined to build rolling-average master frames for each night of data. In the past the ESO/WFI instrument has had problems related to scattered light resulting in the need to apply a zeropoint correction map to the data, as per \cite{2003AJ....126..218M}. With the installation of the baffle in 2006, this has dramatically reduced the scattered light incident on the camera. Attempts to measure the scattered light in WFI in 2012, using a specially designed survey 086.A-9021(A), did not recover a scattered light signature as seen by previous authors. For this reason, we have not applied any correction of this manner. The CASU pipeline was then used to generate aperture photometry and to determine the astrometric solution by cross-matching with the 2MASS point source catalogue. As per MAGIC I, the astrometric solution is generally better than 0$\arcsec$.2. The final photometric catalogue and artificial star tests were undertaken using point-spread function photometry pipeline {\sc dolphot}\footnote{http://purcell.as.arizona.edu/dolphot/} \citep{2000PASP..112.1383D}.  {\sc dolphot} extracts the sources and determines the magnitudes for the two bands, B and R, simultaneously. This was run on all of the fields creating a catalogue of sources per field in both filters. The output files of {\sc dolphot} include detailed information for each star including the position and global solution such as: $\chi$, sharpness, roundness, and object type. For our purposes, we selected  a value for $\chi\geq$ 3.5, reasonable for good stars within relatively crowded fields, a value of (sharpness)$^{2}$$\leq$0.1 for a perfectly-fit star, and object type (which ranges from 1 to 5) of 1 indicating a `good star'. These parameters allow us to remove spurious objects such as extended objects and cosmic rays from the photometry. Typically this corresponds to $\sim$3000 objects removed per field from our area of interest in CMD space.

Artificial star tests were used to determine the photometric completeness of the observations and consisted of $\sim$48300 fake stars per chip. {\sc dolphot} places each artificial star iteratively into the science image and then attempts to extract each them along with the real stars. Each success or failure of this process allows us to build up a profile of the photometric completeness. Following the procedure in MAGIC I, we fitted the ratio of the detected fake stars to total number of fake stars per magnitude bin with a Logistic function and from this we were able to determine the 50\% completeness level of each frame. This information is then utilized in the CMD fitting routine {\sc MATCH} which will be discussed in a later section.

\begin{figure*}
\begin{center}
\includegraphics[width=1\textwidth]{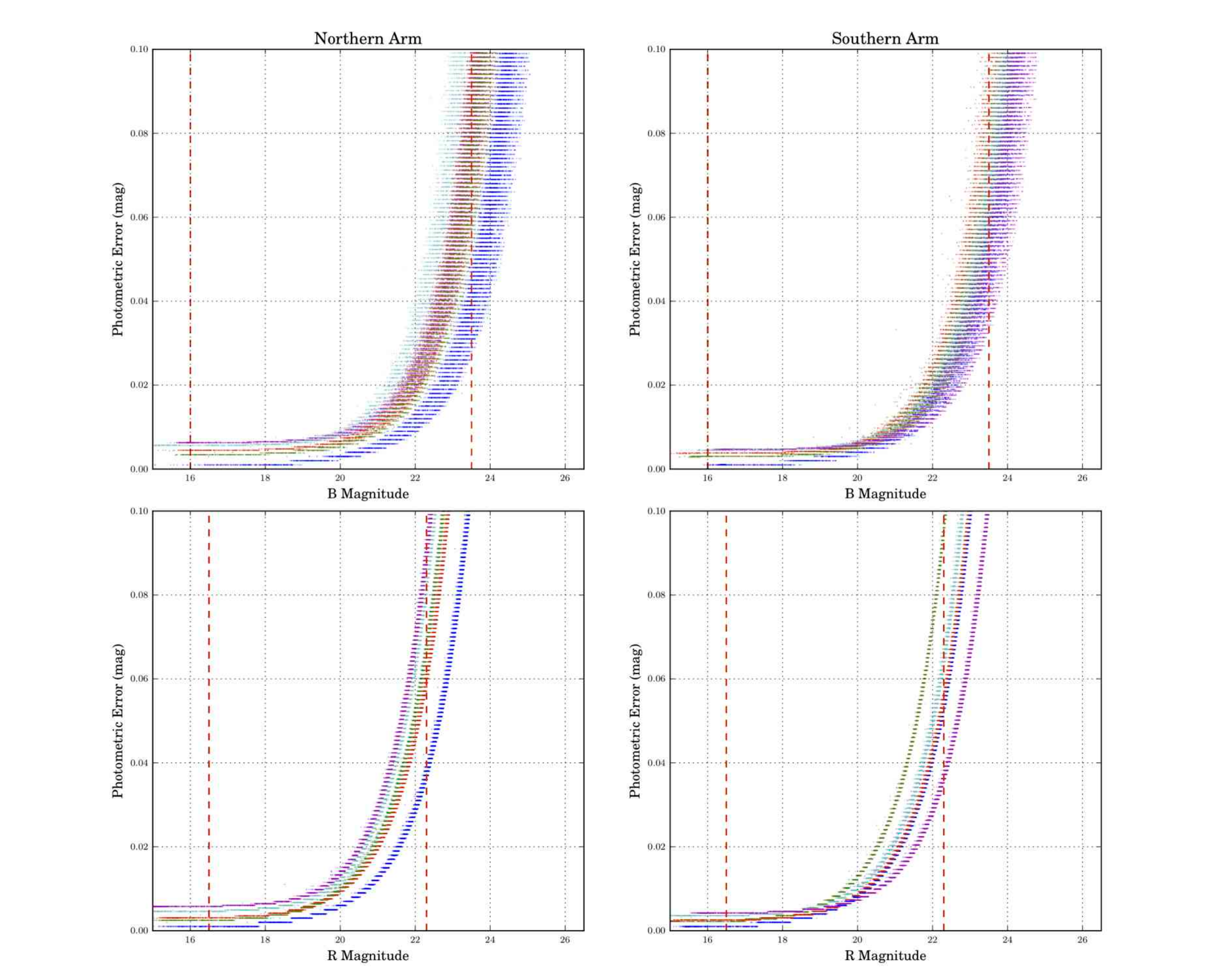}
\caption{Magnitude versus Errors in Magnitude for all fields in MAGIC II. The left panels represent the Northern Arm and the right panels represent the Southern Arm. The top panels are from the B-band photometry and the lower panels from the R-band photometry. The vertical dashed lines delineate the approximate region used in the CMD fitting analysis. The fields (in the bright region) are ordered by increasing error from 1 to 5 in both the North and South. Alternatively the fields 1 through 5 are represented by the colours Blue, Green, Red, Cyan and Magenta.}
\label{errorplot} 
\end{center}
\end{figure*}

\begin{figure*}
\begin{center}
\includegraphics[width=1\textwidth]{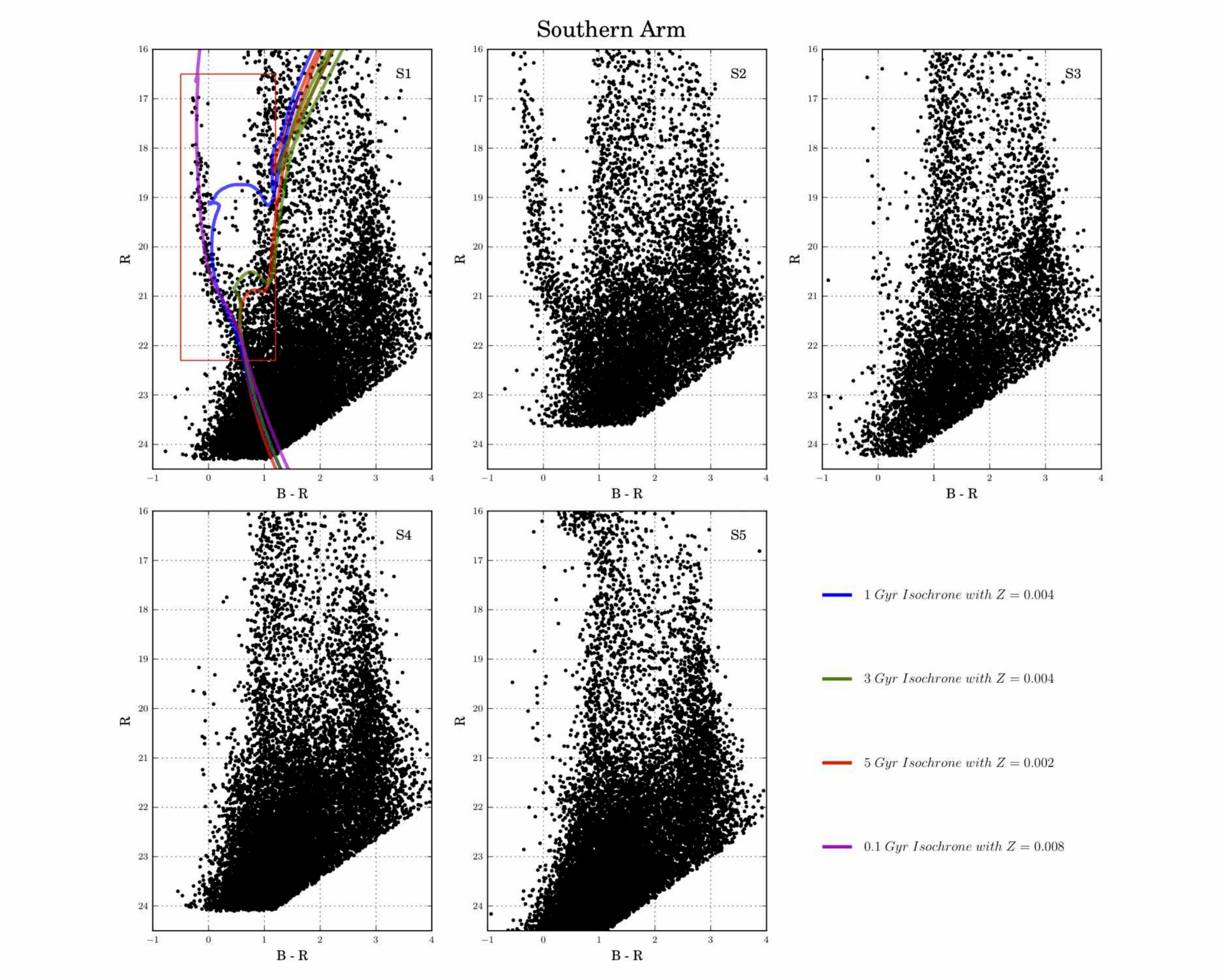}
\caption{CMDs of the Southern arm fields presented in order
of increasing Declination from top left. Red dashed lines represent the region used for the SFH analysis. Isochrones were overlapped showing different evolutionary phases. }
\label{cmds_south} 
\end{center}
\end{figure*}

To calibrate the MAGIC II fields, we use the photometric solution for MAGIC I and from the overlaps between each successive field determine the correction for both the Northern and the Southern Arms. In this process, we tie the instrumental magnitudes of all fields in each arm to the N1 and S1 field respectively, as this only requires a simple offsets between fields. From this solution we then apply the remaining colour terms and offsets determined from the overlap with MAGIC I to correct the photometry to the same Johnson-Cousins system as MAGIC I. We utilised the Markov Chain Monte Carlo code {\sc emcee} \citep{2013PASP..125..306F} to simultaneously calculate the best offsets in both B and R for all fields when determining their relation to N1 and S1. Similarly, when determining the colour term and offset between MAGIC I and N1/S1 we also used {\sc emcee}. In both cases, we typically used over 1000 walkers with more than 5000 steps each. The solutions for the fits can be found in Table~\ref{ObsTable}. The errors in the offsets, as determined using {\sc emcee}, are passed on in quadrature to each successive field. The evolution of the photometric error with magnitude for both the Northern and Southern Arms are shown in Figure~\ref{errorplot}. The spread in errors is greater for the B band as it has a much larger colour term than the R band, which is essentially zero. This has the effect of broadening the error profile. As expected though, at the bright end, the minimum error for each field is greater than the previous field. The resulting CMDs are shown in Figure \ref{cmds_south} and \ref{cmds_north}, where the  CMDs corresponding to the Northern and Southern fields (with respect to B2 of MAGIC I) are presented. The red dashed lines in Figure \ref{cmds_south} show the region used for our analysis here. In order to avoid the contaminants, such as background galaxies, at the faint end of the magnitude limit we only took the stars up to R$=$22.3.

\begin{table*}
\centering
\begin{minipage}{150mm}
\caption{ Summary of Observations. The zero-points of the data are 25.25 in B and 24.68 in R. The atmospheric extinction parameters, $\epsilon_{B}=0.271$ and $\epsilon_{R}=0.07$, were taken from \citep{2006astro.ph.11262G} and \citep{1977Msngr..11....7T}, respectively, as this gave the best B-R colour in comparison with established catalogues. The colour term applied to all fields was 0.223$\pm$ $^{0.013}_{0.001}$ for the B band and -0.013$\pm$ $^{0.005}_{0.001}$}.
\label{ObsTable}
\begin{tabular}{@{}lllllllllllll}
\hline
Field &  RA  & DEC & Date & Seeing  & Filter & Airmass  & 50\% & Mean E(B-V) & Offsets\\
Name & & & Observed & & & & Completeness & &\\
\hline
N1 & 35.668180 & -71.99990 & 2012-01-18 &1.01\arcsec & B\footnote{BB\#B/123\_ESO878} & 1.44 & 25.06 &  0.024& -0.593 $\pm$$^{0.009}_{0.007}$\\
  & 35.667640 & -72.00006 & 2012-01-18 &0.97\arcsec & R\footnote{BB\#Rc/162\_ESO844} & 1.51 & 24.28 &  & -0.724 $\pm$$^{0.012}_{0.01}$\\ 
N2 & 36.166910 & -71.49983 & 2012-01-19 &1.09\arcsec & B & 1.63 & 24.78 & 0.028 & 0.296 $^{ +0.003 }_{ -0.004 }$\\ 
 & 36.166580 & -71.49979 & 2012-01-19 &1.06\arcsec & R & 1.76 & 23.78 &     & 0.119 $^{ +0.004 }_{ -0.001 }$\\ 
N3 & 36.500400 & -71.00002 & 2012-01-20 &1.14\arcsec & B & 1.59 & 24.61 & 0.033 & -0.038 $^{ +0.003 }_{ -0.003 }$\\ 
    & 36.500700 & -70.99997 & 2012-01-20 &1.03\arcsec & R & 1.73 & 23.89 &     &  0.062 $^{ +0.003 }_{ -0.001 }$\\ 
N4\footnote{N4, Chip 7, B offset: -0.5 $^{ +0.003 }_{ -0.003 }$}$^{,}$\footnote{N4, Chip 7, R offset: 0.04 $^{ +0.005 }_{ -0.004 }$} & 37.001040 & -70.49991 & 2012-01-22 &1.42\arcsec & B & 1.40 & 24.67 & 0.036 & -0.01 $^{ +0.003 }_{ -0.003 }$ \\ 
    & 37.000260 & -70.50009 & 2012-01-22 &1.20\arcsec & R & 1.49 & 23.95 &     &  0.004 $^{ +0.003 }_{ -0.004 }$ \\ 
N5 & 37.500900 & -70.00009 & 2012-01-25 &0.91\arcsec & B & 1.41 & 24.89 & 0.043 &  0.101 $^{ +0.003 }_{ -0.003 }$ \\ 
    & 37.500630 & -70.00077 & 2012-01-25 &1.05\arcsec & R & 1.49 & 23.78 &     &  0.082 $^{ +0.004 }_{ -0.003 }$ \\ 
S1 & 35.668100 & -73.50038 & 2012-01-20 &1.07\arcsec & B & 1.55 & 24.67 & 0.040 &  -0.605 $^{0.005}_{0.021}$\\ 
    & 35.667680 & -73.50001 & 2012-01-20 &1.02\arcsec & R & 1.49 & 23.89 &     &  -0.778 $^{0.004}_{0.01}$\\ 
S2 & 36.167360 & -73.99998 & 2012-01-21 &1.38\arcsec & B & 1.64 & 24.39 & 0.043 & -0.049 $^{ +0.003 }_{ -0.003 }$ \\ 
    & 36.168150 & -73.99988 & 2012-01-21 &1.30\arcsec & R & 1.75 & 23.23 &     & -0.068 $^{ +0.002 }_{ -0.002 }$ \\ 
S3 & 36.501480 & -74.49991 & 2012-01-23 &1.65\arcsec & B & 1.71 & 24.00 & 0.043 & -0.051 $^{ +0.002 }_{ -0.003 }$ \\ 
    & 36.500620 & -74.50014 & 2012-01-22 &1.16\arcsec & R & 1.63 & 23.56 &     & -0.111 $^{ +0.001 }_{ -0.001 }$ \\ 
S4 & 37.000630 & -74.99995 & 2012-01-24 &1.14\arcsec & B & 1.51 & 24.84 & 0.036 & -0.158 $^{ +0.004 }_{ -0.001 }$ \\ 
    & 37.001710 & -75.00065 & 2012-01-24 &1.06\arcsec & R & 1.57 & 23.89 &     & -0.021 $^{ +0.002 }_{ -0.003 }$ \\ 
S5 & 37.500590 & -75.49980 & 2012-01-26 &1.05\arcsec & B & 1.52 & 24.84 & 0.037 & -0.104 $^{ +0.002 }_{ -0.001 }$ \\ 
    & 37.498980 & -75.50018 & 2012-01-26 &0.92\arcsec & R & 1.58 & 24.17 &     &  0.005 $^{ +0.001 }_{ -0.003 }$ \\ \hline\end{tabular}
\end{minipage}
\end{table*}

\subsection{CMD fitting}\label{fitting}

The CMD fitting technique was carried out using the software package {\sc MATCH} (\citealt{2002MNRAS.332...91D}) as explained in \cite{2013ApJ...768..109N}  where the reader should refer for more details. In summary, the observed CMDs are converted into Hess diagrams and compared with synthetic CMDs of model populations from \cite{2004A&A...422..205G}. Theoretical isochrones are then convolved with a model of the completeness and photometric accuracy in order to create the synthetic CMDs. We also obtained a foreground estimation based on a Galactic structure model given by {\sc MATCH} that is included in the software as an extra model population. The software uses a maximum-likelihood technique to find the best linear combination of population models, resulting in an estimate of the star formation history (SFH) and the age-metallicity relation (AMR). In order to obtain the SFHs for the Inter-Cloud fields analyzed here, we used the colour range -0.5 $\leq$ (B-R) $\leq$ 1.2 and the magnitude range R $\leq$ 22.3, as marked by the red dashed lines in Figure \ref{cmds_south}. In this way, we avoid the inclusion of red low mass foreground stars in the fitting. 
To further test the influence of unidentified background galaxies contaminating our sample in the region between 21 $\leq$ R $\leq$ 22.3, we calculated the ratio of objects in the red clump to the number of objects at the lower edge of the CMD (21 $\leq$ R $\leq$ 22.3). We find these ratios to be ~10-12\%. This is comparable with the ratio of $\sim$11\% for the wing SMC field from \cite{2007AJ....133.2037N} (qj0111, see grey small diamond in Figure~\ref{fig_bridge}), where the contamination is known to be very low. This, together with the constraints given by {\sc dolphot} described above, assures us that any background contamination has been minimised for our dataset.

We recovered the SFH and metallicity for all the fields using age bins in the range from 10\,Myr to 13\,Gyr and metallicities in the range -2.4$\leq$ [Fe/H] $\leq$ 0 in bins of width 0.2\,dex. We used a Salpeter initial mass function (IMF; since these fields contain predominately intermediate- and old-age stars, the choice of IMF is not critical as shown in \citealt{2010ApJ...714..663D}) and assumed a 30\% binary fraction (see \citealt{2007AJ....133.2037N}). The fitting software then finds a linear combination of model CMDs that best fit the observed CMD. The resulting relative star formation rates (SFRs) and the [Fe/H] as a function of time for the Southern and Northern fields  are presented in Figures \ref{magic2_south} and \ref{magic2_north} respectively.

\section{Results}\label{results}

\subsection{The star formation histories in the MAGIC II fields}

The upper panels of Figures \ref{magic2_south} and \ref{magic2_north} show the SFHs for all fields. The SFR for each bin is the average of the values given by the fits at the different distances and the error bars represent the complete range of SFR values found. As in the case of the MAGIC I fields, we find here a conspicuous old population (older than $\sim$10 Gyr old) present in all fields. 
It is striking to note that the bulk of intermediate-age and young(er) stars increase as we move closer to field B2 of MAGIC I  located in the vicinity of the MB main ridge-line. 
Very little star formation took place in the past 10 Gyr in the Southernmost and Northernmost fields.  These results agree with the proposed numerical scenario where the MB is the result of a recent tidal interaction between the MCs and the MW (\citealt{2007MNRAS.381L..11M}). The amount of HI (as seen from Figure \ref{fig_bridge}) is similar in all the Northern arm fields as in the Southernmost field S5.

The lower panels show the AMRs. The metallicities are only plotted for age bins with significant star formation (SFR $\geq$0.2). The metallicity of each bin is the average of the fits weighted by the SFH. Error bars in the AMR represent the standard deviation. Horizontal bars in both the SFH and the AMR plots represent the width of the age bins used. 
 The AMRs are consistent with being constant on average at [Fe/H]$\sim$-0.8 in agreement with the results of MAGIC I \citep{2013ApJ...768..109N} and with the Calcium triplet metallicities obtained for the outskirts of the SMC (\citealt{2008AJ....136.1039C}) and -- within the quoted uncertainties -- the LMC (\citealt{2011AJ....142...61C}). 
 Recently, \cite{2014MNRAS.442.1680D} found similar results using a factor 10 of more stars.
 Note that these AMRs obtained from CMD fitting should only be taken as indicative of the metallicity range.  

As shown in \cite{2013ApJ...768..109N}, valuable quantitative information on the stellar distribution is given by the cumulative SFH that provides the fraction of stellar mass formed prior to a given time. 
In Figures \ref{cumulative_south} and \ref{cumulative_north} we show these cumulative SFHs for our 10 MAGIC II fields. The upper-$y$-axis shows the redshift as a function of time\footnote{The redshifts were obtained using the following cosmology: $H_{0}$=71, $\Omega$$_{M}$=0.270, $\Omega_\Lambda$=0.730 (http://map.gsfc.nasa.gov/)} with the blue dashed lines mark the time when 50\% of the stars were formed. Although this varies across the fields, most of the fields had 50\% of their stellar population before 10-12 Gyr ago.

The previously introduced Figure \ref{fig_bridge} gives a visual picture that shows the density HI map using data from the LAB survey of Galactic HI \citep{2005A&A...440..775K}. 
The colour contours denote the HI emission levels integrated over the velocity range 80$<$v [km/s] $<$400. Each contour represents the HI column density twice as large as the neighbour contour (Jan Skowron, private communication).
 Our MAGIC II fields are overlapped and depicted as bright magenta, pink and off-white squares (these differences are explained below). MAGIC I field B2 is also shown in Figure \ref{fig_bridge}  as a 
 purple triangle. \cite{2007ApJ...658..345H} fields, \cite{2008MNRAS.389..678B} associations and clusters
in the relevant right ascension (2h-4h), and OGLE IV fields 102 up to 125 from \citealt{2014ApJ...795..108S}
 (without numbers for clarity purposes) are also overlapped. 

\subsection{Southern Arm fields}

Here we concentrate on disentangling the age distribution in the Southern arm fields S1-S5 as seen in Figure \ref{fig_bridge}. 
The MAGIC II fields in bright magenta indicate that we find a fair amount of intermediate-age stars there (between $\sim$27\% and $\sim$48\%). 
Fields in pink show substantial amount of intermediate-age population
 (between $\sim$18\% and $\sim$22\%) while off-white fields indicate that very little intermediate-age population is found there. 
The areas with larger amounts of intermediate-age stars coincide with the higher HI density column but they are more spread out than the HI. 
In these regions, there are a greater amount of young stars, stellar clusters and associations \citep{2008MNRAS.389..678B}. As we move away from the higher HI areas, the amount of intermediate-age population
diminishes until there are very few intermediate-age stars in the Southernmost field S5.
The latter are somehow puzzling since there are young associations found in those parts of the MB. 
 However, our results are in agreement with fields 110, 113 and 114 from \cite{2014ApJ...795..108S} (the OGLE IV fields overlapping our 
 Southern Arm fields) who do not find young or intermediate-age component in those areas. 
Note that the MAGIC I field B2 presents some $\sim$28\% of intermediate-age stars.

From the cumulative SFHs shown in Figure \ref{cumulative_south} we find that fields S1 and S2 show a significant amount of stars formed between 10 and 1\,Gyr ago.
 The remaining stars formed in the last Gyr with a conspicuous increment (20\% of the stars) in the past few Myr. This recent burst is the only clear star burst associated with a pericentric passage between the Clouds 
 ($\sim$\,0.2\,Gyr ago; \citealt{2006ApJ...652.1213K}). 
The arrows in S1 and S2 in Figure \ref{cumulative_south} show the last two pericentric passage between the LMC and the SMC $\leq$2.5 Gyr ago and $\sim$250 Myr ago, respectively, from \cite{2012ApJ...750...36D}. In their best model, \cite{2012ApJ...750...36D}  find that the SMC disk remains intact until the dynamical coupling of the MCs $\sim$2.4 Gyr ago, at which point the Magellanic Stream and the Leading Arm are violently torn away by the strong tidal forces of the LMC. 
They claim that a second tidal encounter occurring $\sim$250 Myr ago is the responsible for pulling the MB from the SMC disc. From our qualitative comparison, there seems to be a correlation between this last pericentric passage between the MCs but no clear connection can be made between the other enhancements in the SFR and the pericentric encounters between the MCs, in agreement with the previous results from \cite{2009ApJ...705.1260N} and MAGIC I.

\subsection{Northern Arm fields}

We focus here in the age distribution in the Northern Arm fields N1-N5 from Figures \ref{magic2_north} and \ref{cumulative_north}. As in the case of the Southern Arm fields, there is a
 bulk of intermediate-age stars in these fields.  
From Figure \ref{cumulative_north} it is clear that almost no stars younger than 1 Gyr old were formed, ruling out the in-situ formation for the Northern Arm. 

The density column of HI in the Northern Arm fields is considerable lower than in the Southern Arm. However we still see intermediate-age stars indicating that the young stars follow the HI gas but the
intermediate-age population is considerably more spread out.

\subsection{Disentangling the Age Spread}  

\begin{figure*}
\begin{center}
\includegraphics[width=1\textwidth]{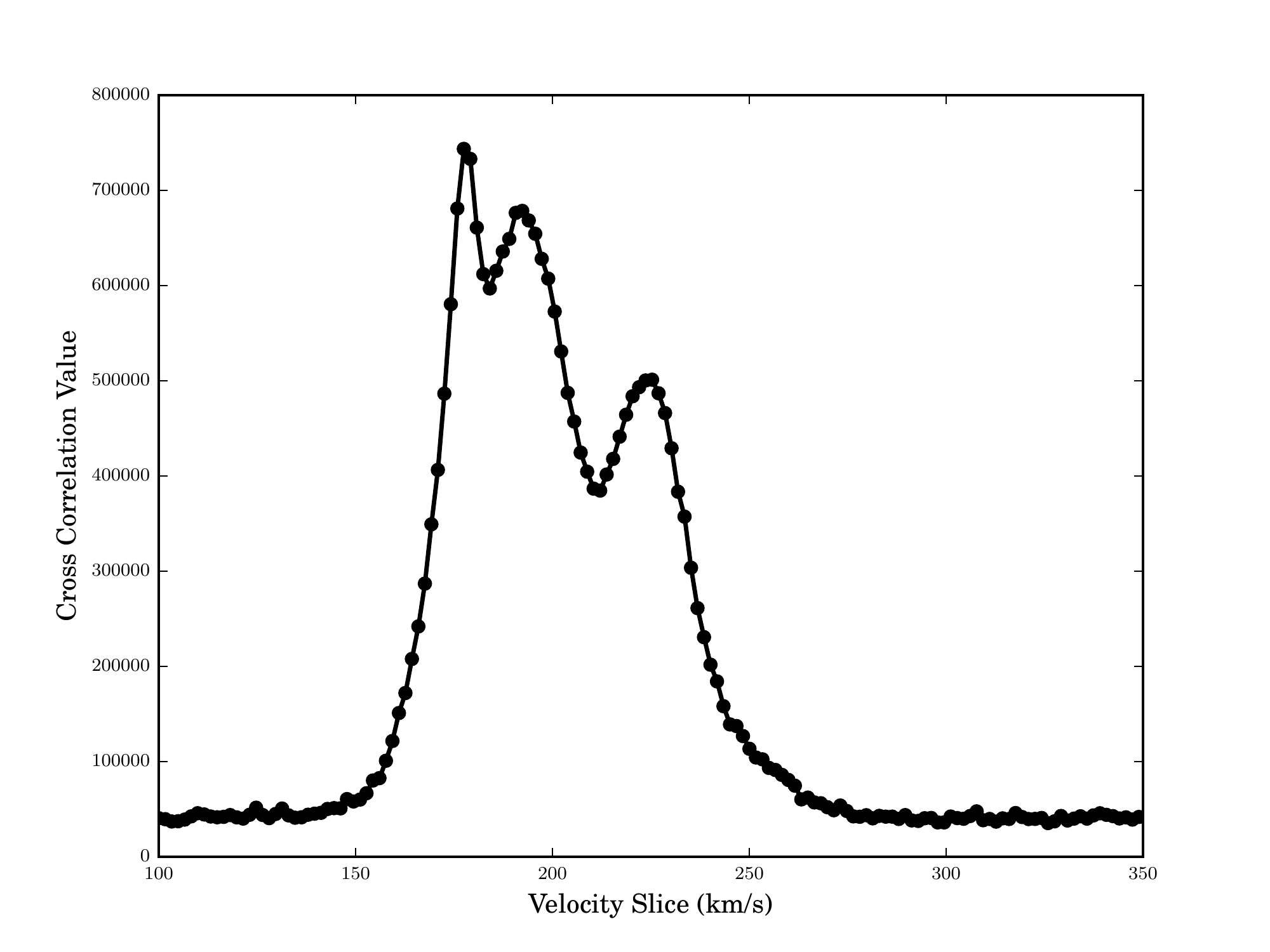}
\caption{\label{yms}
Spatial distribution of the MAGIC II young MS stars are binned into the same spatial grid as the \citealt{2003MNRAS.339..105M} gas maps. 
The x-axis represents the different velocity slices as per the \citealt{2003MNRAS.339..105M} maps and the y-axis is the sum of the stellar density map multiplied by the gas density at that velocity. 
If the gas follows stars then the highest correlation between the maps occurs at 
$\sim$180 Km/s, while there are two other velocity peaks at $\sim$193 and $\sim$225 Km/s.}
\end{center}
\end{figure*}

We now turn our attention to the correlation between young, intermediate and old-age stars in the MAGIC II fields and the HI gas. 
Without velocity information, we cannot determine conclusively if such correlations exist, but marginalising over the unknown stellar velocities we can ask if such correlations are at least supported by the data. In Figure            \ref{yms}, we look for evidence for a correlation between the HI gas and young stars in our fields.
 We sum up the number of young MS stars in each pixel of the SMC gas map from \cite{2003MNRAS.339..105M}. Stepping through each velocity slice of these radio data, we multiplied the star and gas densities and summed all positive pixels. The sum of this convolution as a function of the velocity slice is shown in Figure \ref{yms}. 
We find three velocity peaks where the stars and gas density correlate strongly, suggesting that it is at least possible for the young stars and HI gas to be highly correlated. 

Unfortunately, we cannot repeat the above analysis for our intermediate age and old-stars since these cannot be so easily identified by their colour and magnitude alone. Instead, we consider how the intermediate-age and old stars relate spatially to our young star population at the spatial resolution of each MAGIC-II field. This is shown in Figure \ref{three_pops}. The greyscale contour shows the percentage of young (left), intermediate (middle) and old-age (right) stars in each of our MAGIC II fields. These are overlaid on gas density contours taken from \cite{2014ApJ...795..108S}  as in Figure \ref{fig_bridge}.

Notice that the young stars are clustered strongly around our MAGIC I field B2 (marked by the purple triangle in Figure \ref{fig_bridge}). By contrast, the intermediate-age stars are more extended to the north and south, while the old stars are spread with almost constant density across all fields. Thus, if the young stars correlate strongly with the HI gas then the intermediate and old-age stars do not. We discuss the implications of this in section \ref{discussion}.


\subsection{Surface Brightness Profiles}

We derived the surface brightness for the ten fields analyzed here using the selection box shown in Figure~\ref{cmds_south} as red dashed lines. The surface brightness is measured in B-band as a function of distance from the mass centre of the SMC. 
We integrated the flux in the CMD subtracting the flux contribution from foreground MW halo giants and metal poor MW dwarfs and from background galaxies. 
In order to estimate and remove the contamination, we used the predictions from the TRILEGAL Galaxy model \citep{2005A&A...436..895G}
scaling the number of predicted MW stars to the number observed in our fields in those parts of the CMDs that were clearly 
devoid of MB Inter-Cloud stars (see \citealt{2004ApJ...614L.109G} and \citealt{2007ApJ...665L..23N} for more details).

We obtained two separate fluxes: the total flux from stars older than 10 Gyr old (considered the old population)  and the total flux from intermediate-age  stars (between 1 and 10 Gyr old).  The results are
shown in Figure \ref{surfacebr} where the squares represent the Southern fields and the circles depict the Northern fields. Open (red in the colour version) symbols denote old populations while
filled (green in the colour version) ones denote the intermediate-age populations.  Error bars represent the difference between the values of the
foreground flux in the fiducial CMD area as predicted by the TRILEGAL code and after scaling by the ratio of observed and
predicted fluxes (see \citealt{2007ApJ...665L..23N}).
As seen from the figure, both profiles (from the old and from the intermediate-age populations) are relatively flat within the errors instead of the expected exponential at inner radii from the SMC centre. 
This is in agreement with the density profile from \cite{2011ApJ...733L..10N} who found that at these remote distances the star counts profile deviates from 
the exponential fit, showing a ``break'' in the population that extends up to 10.6$^{o}$ from the SMC center.
\cite{2011ApJ...733L..10N} argue that the origin of this break population is either a tidal debris or a ``classical'' bound stellar halo similar to the one around the MW  and M31. 
We cannot, however, directly compare their analysis with ours since \cite{2011ApJ...733L..10N} deliberately removed their outer fields in the direction toward the LMC.  

The most plausible explanation is that the close pericentre passage between the LMC and the SMC $\sim$200 Myr ago (e.g. \citealt{2007ApJ...668..949B}; 
\citealt{2009ApJ...705.1260N}), believed to have stripped HI to form the MB, could have also tidally stripped stars of all ages from the SMC as indicated by the similar spatial distribution of the old and intermediate-age profiles in Figure \ref{surfacebr}. 
The fact that intermediate-age stars show a brighter surface brightness profile could mean that the number of these intermediate-age stars is larger than the old one
 or that the intermediate-age stars are brighter than the old stars.

Finally, it is important to note that the intermediate-age and old stars in Figure \ref{surfacebr} are similarly distributed, while in Figure \ref{three_pops} they have clearly different spatial distributions. This apparent contradiction arises because in Figure \ref{three_pops} we see that the young, intermediate and old age stars are distributed differently with respect to the HI gas in the Magellanic Bridge. By contrast, in Figure \ref{surfacebr}
 we consider instead how they are distributed with respect to distance from the SMC. Over the small radial range of 6.5 to 8.5 Kpc from the SMC centre probed by our MAGIC II fields, we are unable to detect a 
 difference in the radial profile of the intermediate and old-age populations (within our quoted uncertainties). It is likely, however, that over large radial ranges and/or with more complete spatial coverage we would see 
 rather different distributions for the intermediate-age and old-age stars, as we do in Figure \ref{surfacebr}.

\section{Discussion and Conclusions}\label{discussion}
 
We have presented the stellar content and distribution of a transversal strip perpendicular to the main HI ridge of the `Inter-Cloud' region. 
The fields were chosen deliberately to allow us to trace the stellar populations of the Inter-Cloud radially out from the SMC since thus the contaminating halo Galactic stars can be removed and allowing us to probe the
 outskirts of the SMC in the MB direction. 
 
We confirm here the results of MAGIC I, finding that the stellar populations and mean metallicities of the fields are similar to fields to the South and West of the SMC but at smaller radii from the SMC center
  ( $\sim 2.5$\,Kpc as in the case of \citealt{2009ApJ...705.1260N}). 
  The similarity between the MAGIC I and MAGIC II fields to some SMC fields that lie much closer in projection to the SMC is an indication that these MB fields contain tidally stripped SMC stars.

\begin{figure*}
\begin{center} 
\includegraphics[width=1\textwidth]{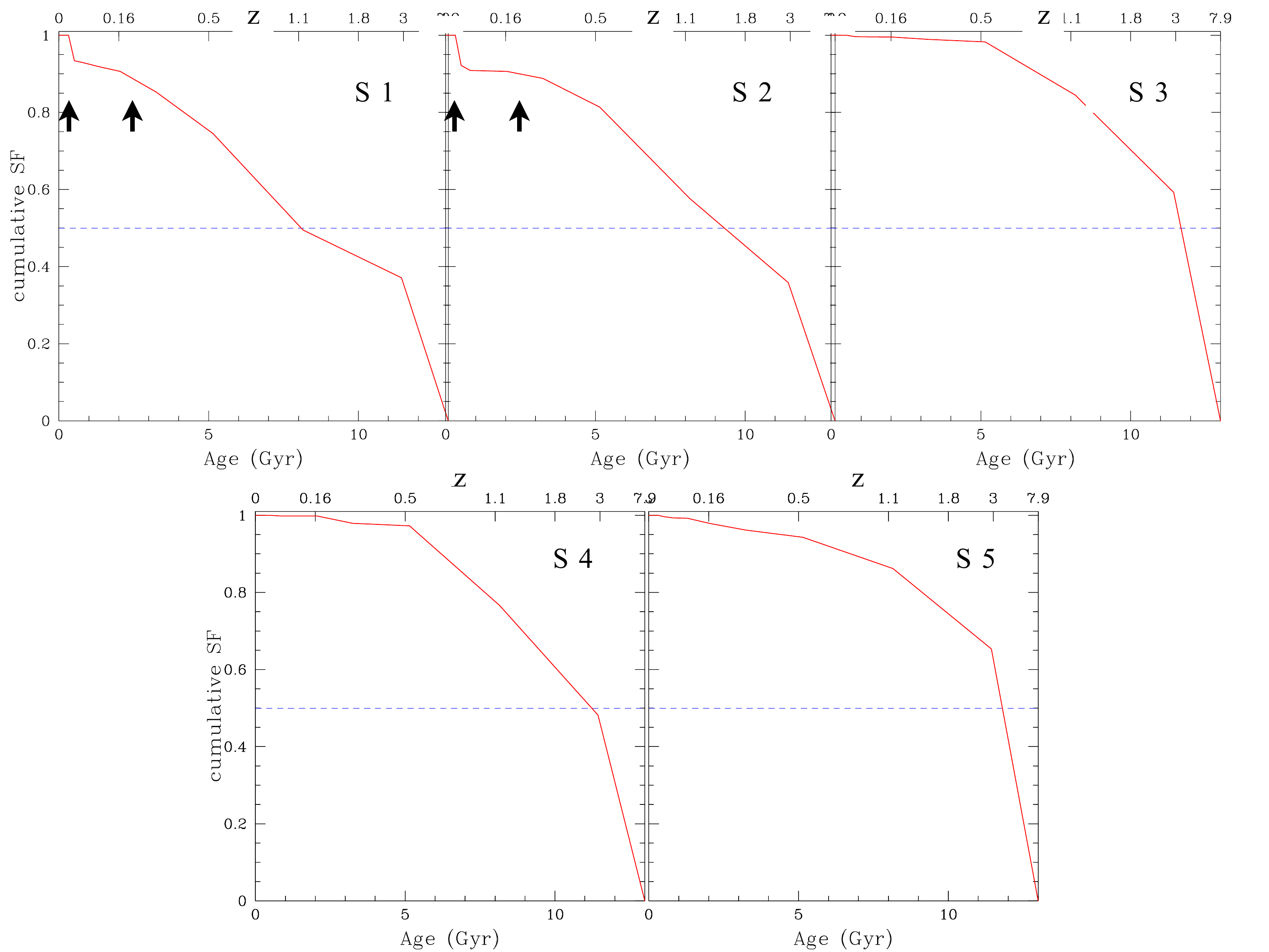}
\caption{\label{cumulative_south}  Cumulative SFHs. The blue dashed horizontal line shows when 50\% of the stars were formed. The upper $y$ axis shows redshift for our current cosmology. The (tentative) pericentre passages that indicate possible past interactions between the LMC and SMC as taken from \citealt{2012ApJ...750...36D} are marked by arrows in S1 and S2.}
\end{center}
\end{figure*}

In good agreement with MAGIC I and previous studies, we found significant intermediate-age stars in the bridge region and young stars of a similar age to the last pericentre passage 
in-between the clouds $\sim$200-250 Myr ago, as determined from their proper motions and radial velocities  (see Figure ~\ref{cumulative_south}). 
These recent enhancements are likely associated with the recent pericentric passage between the clouds (\citealt{2006ApJ...652.1213K}).
However, it is interesting to note that we find no such prominent bump in the star formation at earlier times. This could owe to the difficulty of determining orbits backwards in time
 \citep{2010MNRAS.406.2312L}, or the poorer temporal resolution of the SFH backwards in time \citep{2012ApJ...751...60D},  or could indicate that the two clouds have only just interacted for the very first time 
 \citep{2012ApJ...750...36D}.

\begin{figure*}
\begin{center}
\includegraphics[width=1\textwidth]{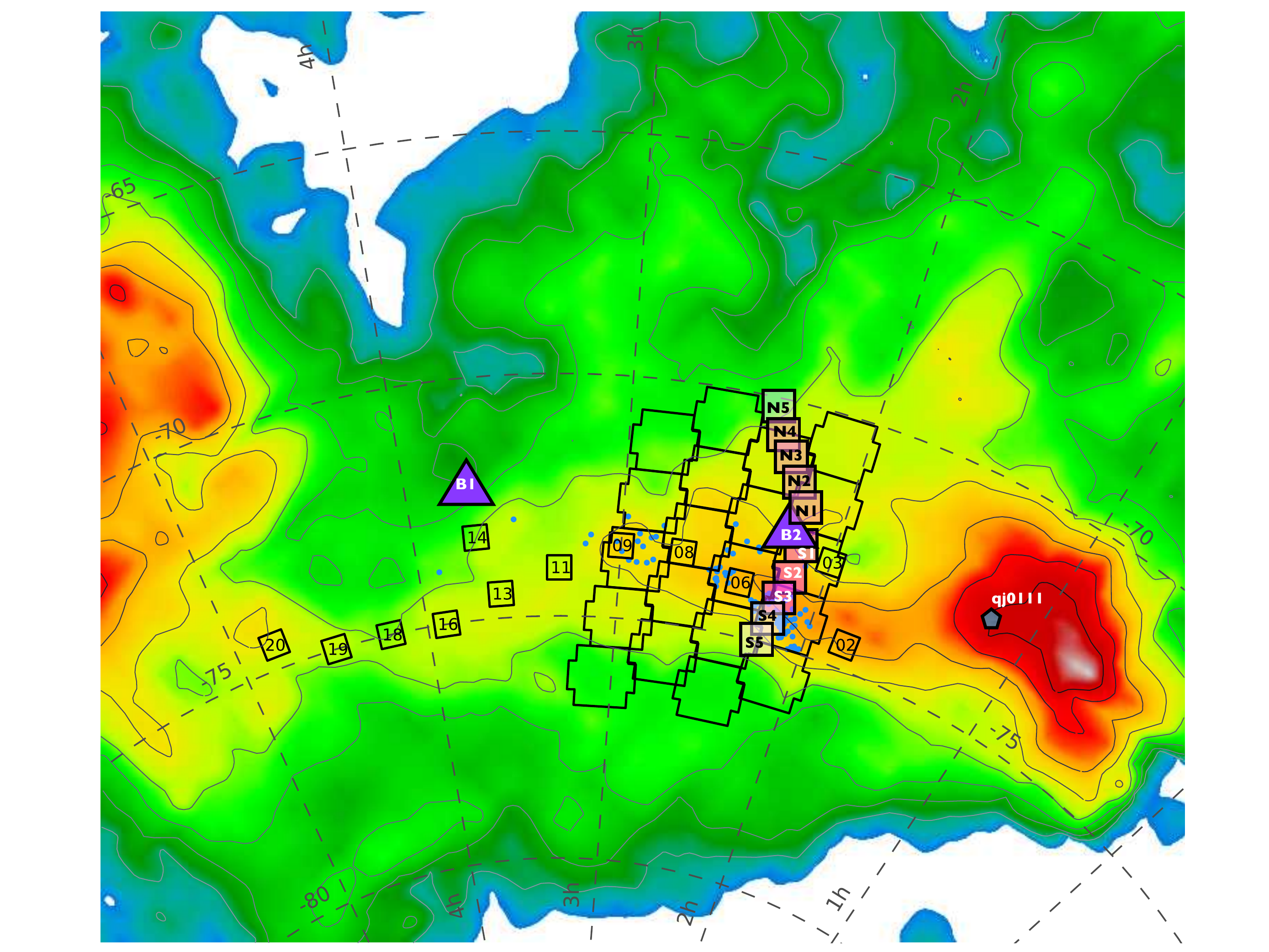}
\caption{\label{fig_bridge} Spatial location of the Inter-Cloud fields presented in this paper represented by squares, together with the two fields from MAGIC I, B1 and B2, depicted
as the two purple triangles. Fields from \citealt{2007ApJ...658..345H} are also overlapped with the corresponding numbers as well as OGLE IV fields 106 up to 125 from \citealt{2014ApJ...795..108S} (we avoided the numbers here for clarity purposes). \citealt{2008MNRAS.389..678B} clusters and associations in the area around our MAGIC II fields are represented with blue dots. The different shades in the colours of the MAGIC II fields describe the amount of intermediate-age stars present in each field: the off-white squares depict fields with little  intermediate-age population, the light pink squares represent those areas where there is substantial intermediate-age stars, and the bright magenta squares where there is significant intermediate-age population.  The colour contours represent the HI emission integrated over the velocity range 80$<$ v [km/s] $<$400, where each contour represent the HI density column taken from the Leiden/Argentine/Bonn -LAB- survey of Galactic HI   \citep{2005A&A...440..775K};  see Figure 8 from \citealt{2014ApJ...795..108S} for more details.}
 \end{center}
 \end{figure*}

\begin{figure*}
\begin{center}
\includegraphics[width=1\textwidth]{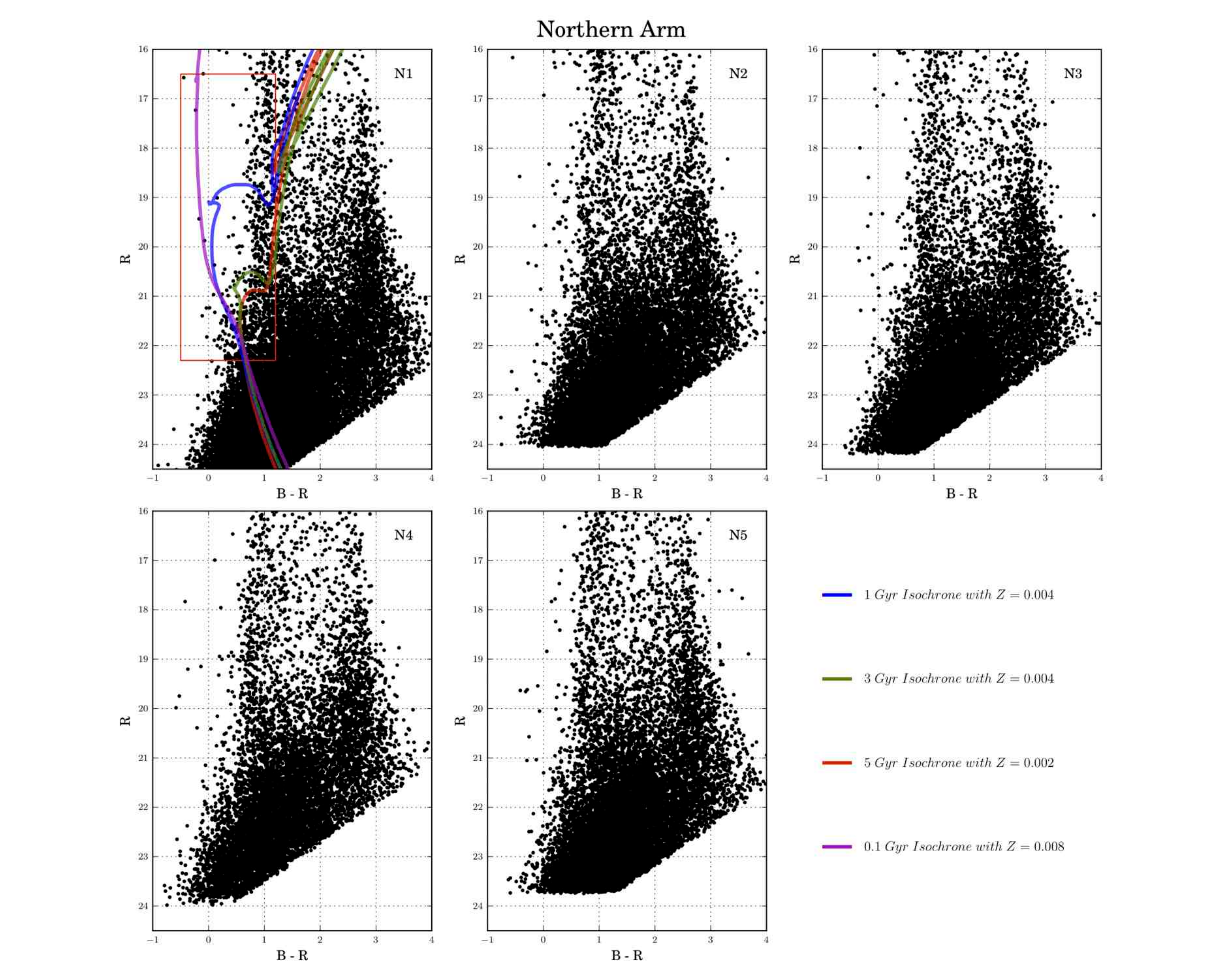}
\caption{CMDs of the Northern arm fields presented in order
of decreasing Declination from top left. Red dashed lines represent the region used for the SFH analysis. Isochrones were overlapped showing different evolutionary phases (see figure~\ref{fig_bridge} for a visual clarification).}
\label{cmds_north} 
\end{center}
\end{figure*}

\begin{figure*}
\begin{center}
\includegraphics[width=1\textwidth]{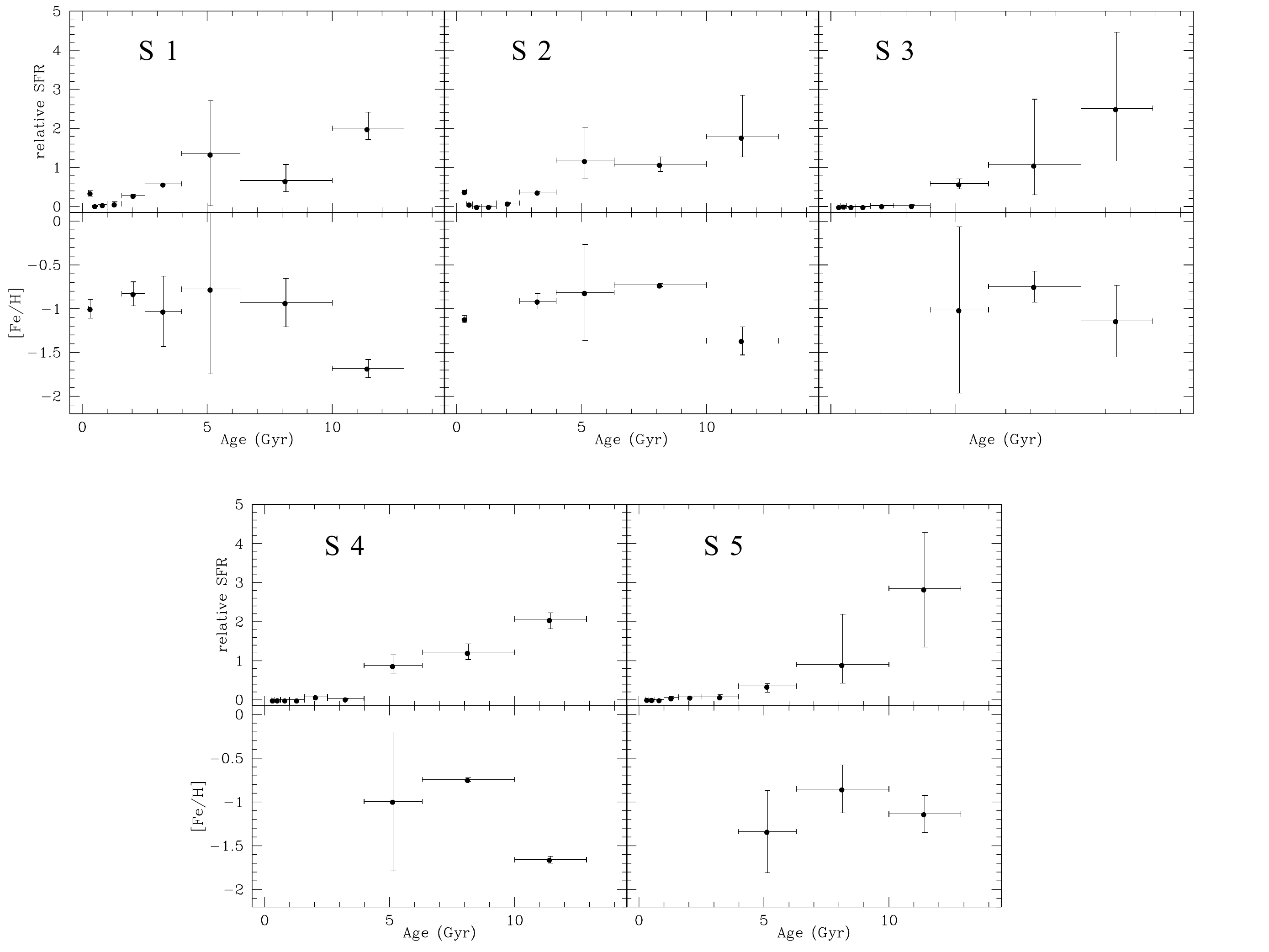}
\caption{\label{magic2_south} SFH and AMR for the five Southern MAGIC II fields. Clockwise from top left panel in order of increasing Declination SFHs and AMRs for the ten MAGIC II fields (see figure~\ref{fig_bridge}). Each Figure shows the average SFRs over the total age range (upper panel) and 
the SFR-weighted metallicity, i.e. the AMR (lower panel). The vertical error bars in the upper panels give the full range of SFRs found in the fits while the vertical error bars in the lower panels give the standard deviation in the metallicities of the fits. Horizontal bars in both the upper and lower panels indicate the width of the age bins.}

\end{center}
\end{figure*}

\begin{figure*}
\begin{center}
\includegraphics[width=1\textwidth]{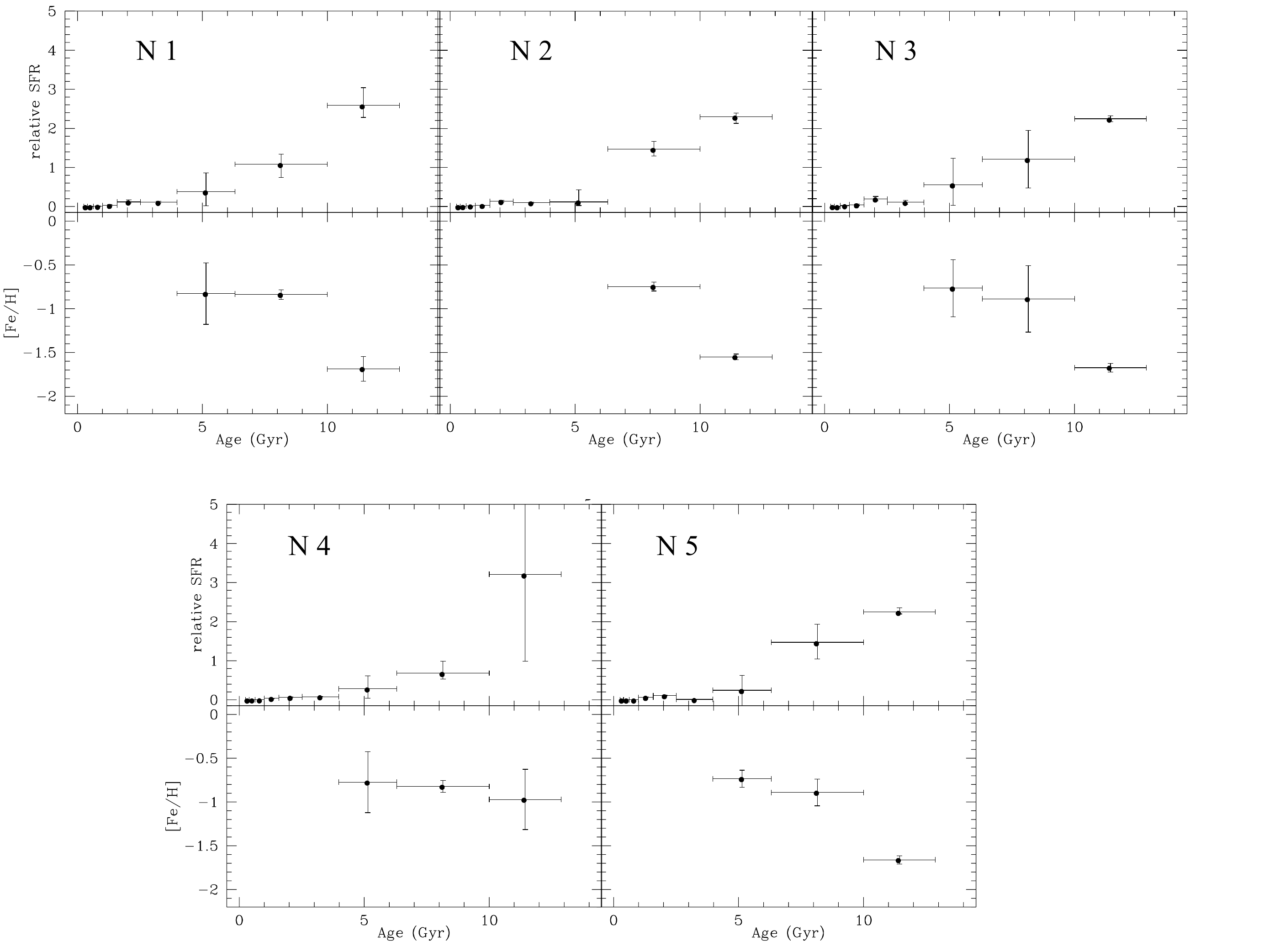}
\caption{Same as figure~\ref{magic2_south} for the fields North of B2 in MAGIC I (see figure~\ref{fig_bridge} for a visual clarification).}
\label{magic2_north} 
\end{center}
\end{figure*}

\begin{figure*}
\begin{center}
\includegraphics[width=1\textwidth]{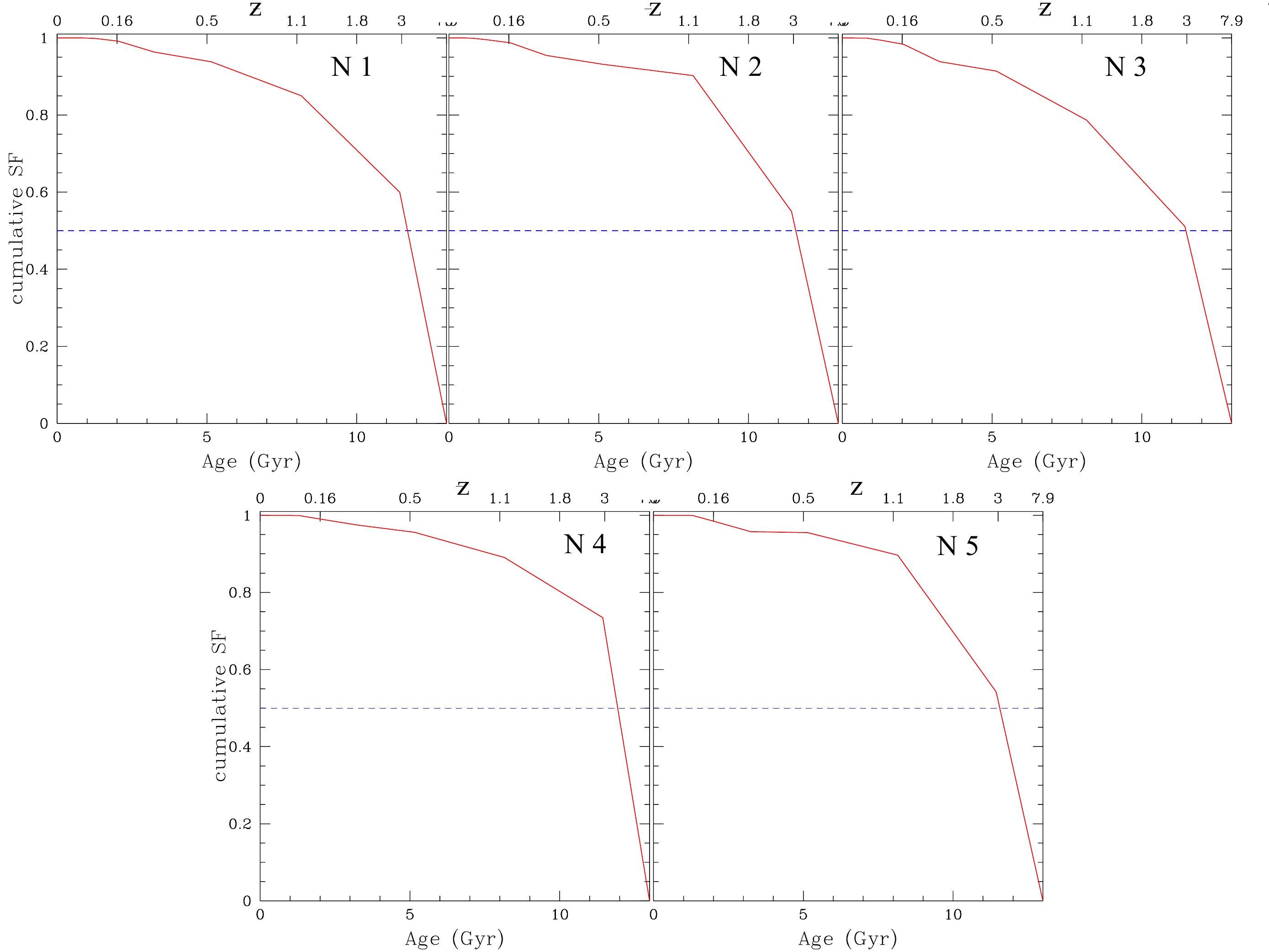}
\caption{\label{cumulative_north} Same as figure~\ref{cumulative_south} for the fields in the Northern arm (see figure~\ref{fig_bridge} for a visual clarification).}
\end{center}
\end{figure*}

\begin{figure*}
\begin{center}
\includegraphics[width=1\textwidth]{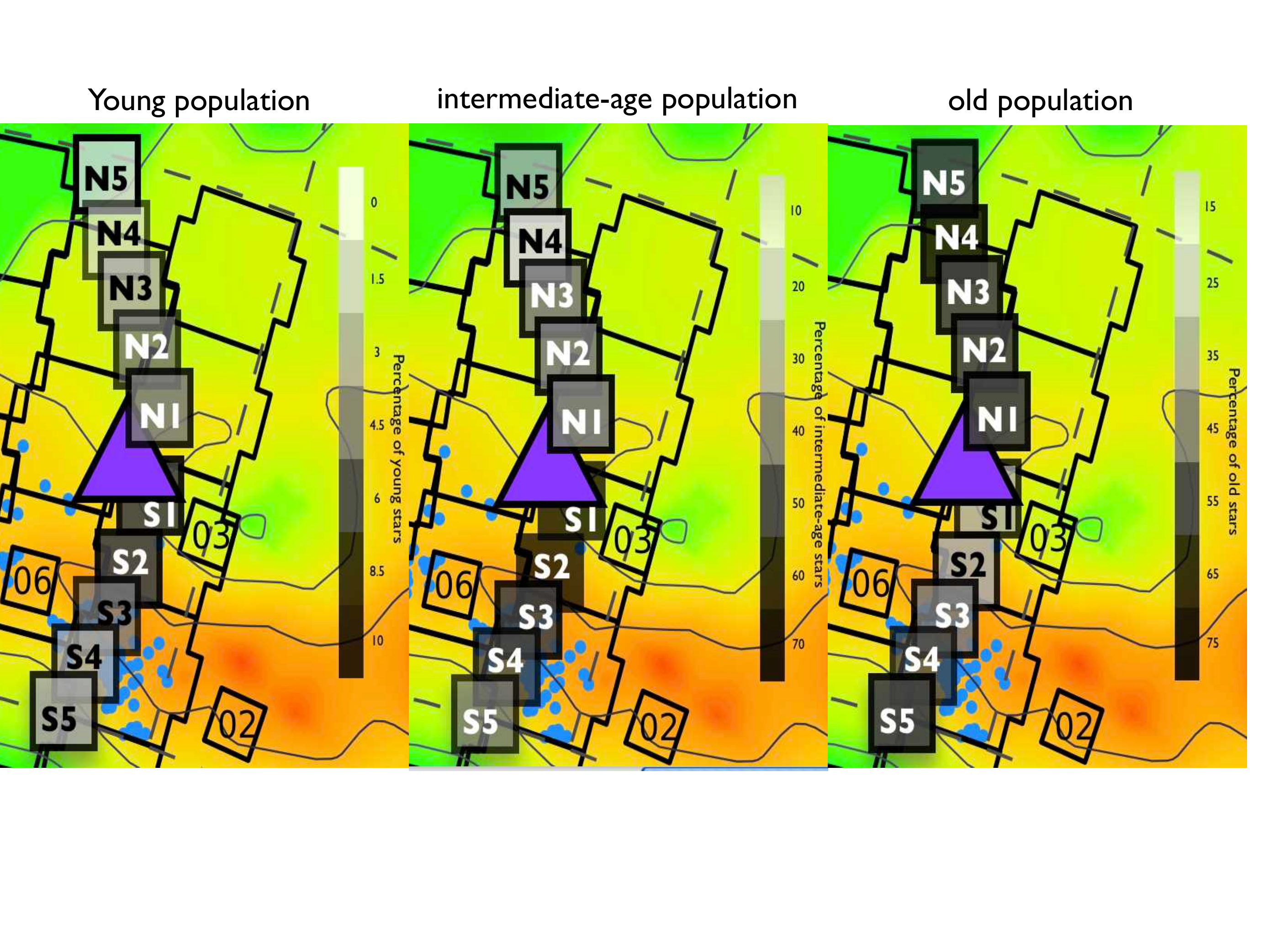}
\caption{\label{three_pops} The percentage of young (left), intermediate-age (middle), and old (right) stars for our MAGIC II fields (grey contours). The coloured contours show the HI gas map from Figure \ref{fig_bridge}.
}
\end{center}
\end{figure*}

\begin{figure*} 
\begin{center}
\includegraphics[width=1\textwidth]{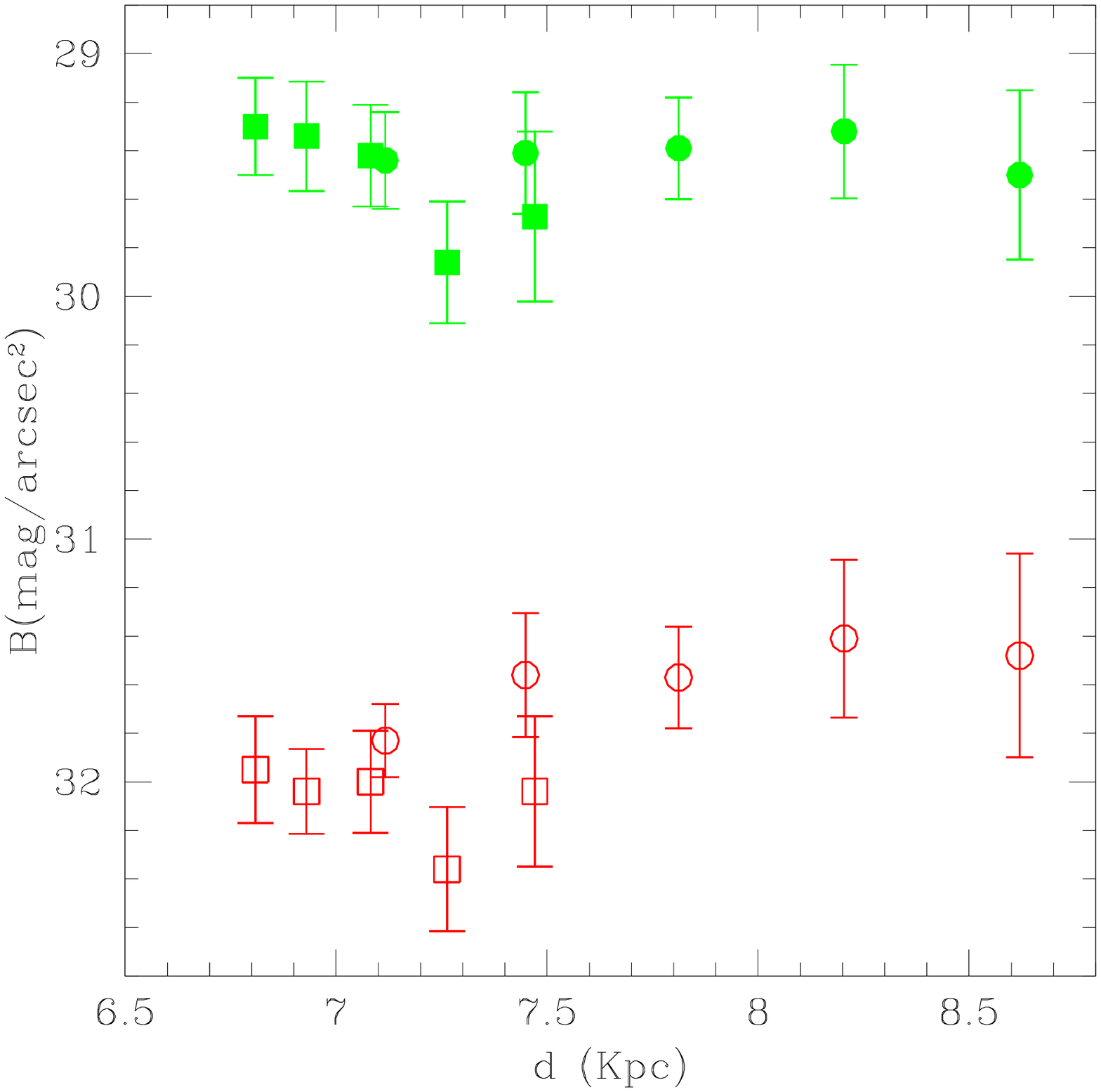}
\caption{\label{surfacebr}  B-band surface brightness profile of the MAGIC II fields presented here. 
The intermediate-age (filled green symbols) and old stars (open red symbols) in each field were split in order 
to quantify the contribution of the different populations. Circles represent the Northern Arm fields
while squares denote the Southern Arm fields. See figure~\ref{fig_bridge} for a visual location of the fields.}
\end{center}
\end{figure*}

We showed here for the first time that the young, intermediate-age and old stars have distinct spatial distributions with respect to the HI gas in the MB. Our key results are as follows: 

\begin{enumerate}

\item The young stars correlate well with the HI gas suggesting that they were either recently stripped from the SMC or formed in-situ. 
Unfortunately, at our current CMD-age resolution these two options are degenerate. Based on their proper motions, the clouds likely interacted just $\sim$200 Myr ago \citep{2012MNRAS.421.2109B}
 which is comparable to the recent star formation burst (see Figure \ref{cumulative_south}). Thus, these stars could have formed in the body of the SMC and had time to be tidally torn into the MB 
 (along with the bridge gas); or they could have formed in-situ. We note that both possibilities could act in tandem, while both lead to young stars that are highly correlated with dense gas.

\item The intermediate-age stars (1-10 Gyr old) are located mainly in the bridge region but are more spread out than the young stars, as seen in Figure \ref{fig_bridge}. 

\item The intermediate-age population has very similar properties to stars located $\sim$2 Kpc from the SMC centre, suggesting they were tidally stripped from this region. 

\item Finally, the old stars extend to some $\sim$9 Kpc from the SMC supporting the idea that all galaxies have a large extended metal poor stellar halo. 

\end{enumerate}

The work presented here and in MAGIC I represent a complex picture for the MB region. There appears to be young stars that correlate with HI gas;
a spatially distinct intermediate-age component; and a possibly smoother old-age component. It is important to emphasize that this study shows the power of deep resolved CMDs as a tool for  empirically unravelling the past history of even rather complex galactic systems like the Magellanic Clouds.

In a future paper, we will use stellar kinematics in order to pin down whether the gas and young stars perfectly correlate; to explore the role of ram-pressure on the gas; to test whether the young stars 
formed in-situ; and to further probe the likely tidally-stripped intermediate-age stars.

\section*{Acknowledgments}

We thank the anonymous referee for their helpful comments that enhanced the manuscript.
We would like to thank Jan Skowron for his invaluable help in producing Figure~\ref{fig_bridge}. We also thank  O. Agertz, C. Gallart, N. Martin and M. Monelli for useful discussions. 
NEDN would like to thank the Department of Physics at the University of Surrey. BCC  thanks Gemini Observatory and the Max Planck Institute for Astronomy and the Alexander von Humbodlt Foundation for their support during this project. 
JIR would like to acknowledge support from STFC consolidated grant ST/M000990/1 and the MERAC foundation. RC acknowledges financial support provided by the Spanish Ministry of Economy and Competitiveness under grant AYA2010-16717.

\end{document}